# Phonons and electron-phonon coupling in the phonon-mediated superconductor YNi$_2$B$_2$C

F. Weber[1], L. Pintschovius[1], W. Reichardt[1], R. Heid[1], K.-P. Bohnen[1], A. Kreyssig[3,4], D. Reznik[5,6], K. Hradil[7a]

[1] *Institut für Festkörperphysik, Karlsruher Institut für Technologie, P.O. 3640, D-76021 Karlsruhe, Germany*
[3] *Institut für Festkörperphysik, Technische Universität Dresden, D-01062 Dresden*
[4] *Ames Laboratory, Iowa State University, Ames, IA-50011, USA*
[5] *Laboratoire Léon Brillouin, CEA Saclay, F-91911 Gif-sur-Yvette, France*
[6] *Department of Physics, University of Colorado at Boulder, Boulder, Colorado 80309, USA*
[7] *Universität Göttingen, Institut für physikalische Chemie, Außenstelle FRM-II, Lichtenbergstr. 1, D-85747 Garching*

**We present a combined density-functional-perturbation-theory and inelastic neutron scattering study of the lattice dynamical properties of YNi$_2$B$_2$C. In general, very good agreement was found between theory and experiment for both phonon energies and line widths. Our analysis reveals that the strong coupling of certain low energy modes is linked to the presence of large displacements of the light atoms, i.e. B and C, which is unusual in view of the rather low phonon energies. Specific modes exhibiting a strong coupling to the electronic quasiparticles were investigated as a function of temperature. Their energies and line widths showed marked changes on cooling from room temperature to just above the superconducting transition at T$_c$ = 15.2 K. Calculations simulating the effects of temperature allow to model the observed temperature dependence qualitatively.**

---

[a] Current address: Röntgenzentrum, Technische Universität Wien, A-1060 Wien, Austria

## I. Introduction

On cooling below the superconducting transition temperature $T_c$, Cooper pairs are formed where two electrons are bound to each other by an attractive interaction (the "glue") [1]. For many superconductors (termed "conventional"), the pairing mechanism is based on electron-phonon coupling (EPC). Although it is known that phonons have a momentum and energy dependent coupling strength to electronic states near the Fermi energy $E_F$, the interaction is often cast into a single constant $\lambda$, which represents the EPC strength of individual phonon modes integrated over all phonon branches and the whole Brillouin zone.

Inelastic neutron scattering is, in principle, able to provide all the necessary information to experimentally obtain $\lambda$, measuring all phonon energies $\hbar\omega_q$ and also line widths $\gamma_q$, which are strongly influenced by EPC [2]. In reality, neutron scattering investigations of superconductors were typically restricted to phonons with energies close to the superconducting energy gap $2\Delta$ and focused on temperature dependencies across the superconducting transition temperature [3,4]. A prominent exemption is the combined theoretical and experimental investigation of phonon line widths in the elemental superconductor Nb in the late 1970s [5,6].

Modern *ab-initio* calculations based on *density-functional-perturbation-theory* (DFPT) predict the properties of electrons, phonons and their coupling in great detail [7]. However, results are often only interpreted in terms of $\lambda$ and the calculated superconducting transition temperature $T_c$. In order to perform a more detailed check of current calculations we embarked on a comprehensive investigation of the phonon properties of $YNi_2B_2C$ using both DFPT calculations and inelastic neutron scattering. We note that we have chosen $YNi_2B_2C$ ($T_c$ = 15.2 K) and not $YPd_2B_2C$ ($T_c$=23K [8]) – which has a higher $T_c$ - for our study for the simple reason that there are no single crystals available for $YPd_2B_2C$. $YNi_2B_2C$ has been already investigated by two other groups [4,30] but only with a very limited scope, i.e. to study the temperature dependence of a a pronounced phonon anomaly in a transverse acoustic branch across the superconducting transition. Our study had a much broader scope, aiming at phonon energies of all phonon branches throughout the Brillouin zone and moreover, great care was taken to determine phonon line widths as well.

We note that the neutron scattering technique has to be pushed to its limits to achieve a sufficiently high resolution to determine the intrinsic phonon line widths related to EPC, because these intrinsic line widths rarely exceed a few percent. We focused on those phonon branches where theory predicted line width values detectable by experiment. Further, we tried to check theoretical predictions as to the phonon eigenvectors in those cases where density functional theory and a simple force constant model gave very different results.

Phonons having a particularly strong EPC were investigated not only at very low temperature, but also at elevated temperatures up to room temperature, because substantial changes of energies and/or line widths were found in this temperature range. This observation motivated additional calculations trying to understand the strong temperature dependence of these phonons. To this end, the smearing of the electronic energy levels used in the calculations to facilitate numerical convergence was varied on purpose over a wide range. As we will show, such an approach allows one to model the observed temperature dependence of the phonon properties in a qualitative way, as it mimics the smearing of the Fermi edge with temperature.

On cooling below the superconducting transition temperature $T_c$, some low energy phonons show drastic changes of their line shape. These changes, which can be understood by a different theory focused specifically on this phenomenon, are outside the scope of this paper. They were already discussed in a separate publication [9]. Preliminary theoretical and experimental results of our study were published in conference proceedings [10]. More recently, results for the high energy modes above 70 meV were published in [11].

## II. Theory

The calculations reported in this paper were performed in the framework of density functional theory with the mixed basis pseudopotential method [12]. Scalar-relativistic norm-conserving pseudopotentials were constructed following the descriptions of Hamann-Schlüter-Chiang [13] for Y, and of Vanderbilt [14] for Ni, C, and B. In the case of Y, the semicore *4s* state was treated explicitly as valence state. The mixed-basis scheme uses a combination of local functions and plane waves for the representation of the valence states [12], which allows for an efficient treatment of the fairly deep norm-conserving pseudopotentials. The local basis was supplemented by plane waves up to a kinetic energy of 26 Ry. For the exchange correlation functional, the local-density approximation (LDA) in the parameterization of Hedin-Lundqvist [15] was applied.

Phonon properties and EPC matrices were calculated using the linear response technique or density functional perturbation theory (DFPT) [7] in combination with the mixed-basis pseudopotential method [16].

Brillouin zone (BZ) integrations were performed by **k**-point sampling in conjunction with the standard smearing technique [17] employing a Gaussian broadening of 0.02 – 0.2 eV. The **k** point meshes were adapted to the simple tetragonal cell following the scheme of Moreno-Soler [18]. To obtain sufficient convergence, especially in the neighborhood of anomalies, a mesh corresponding to 207 **k**-points in the irreducible BZ (IBZ) was used for phonon calculations as well as for structural optimization.

An even denser mesh with 1309 **k**-points in the IBZ was used for the calculations of EPC matrix elements, which involve slowly converging Fermi-surface averages.

## III. Experiment

The neutron scattering experiments were performed on the 1T triple-axis spectrometer at the ORPHEE reactor at LLB, Saclay, and on the PUMA triple-axis spectrometer at the research reactor FRM II in Munich. Double focusing pyrolytic graphite monochromators and analyzers were employed in both cases for phonons with energies below 20 meV. High energy phonons were measured with Cu220

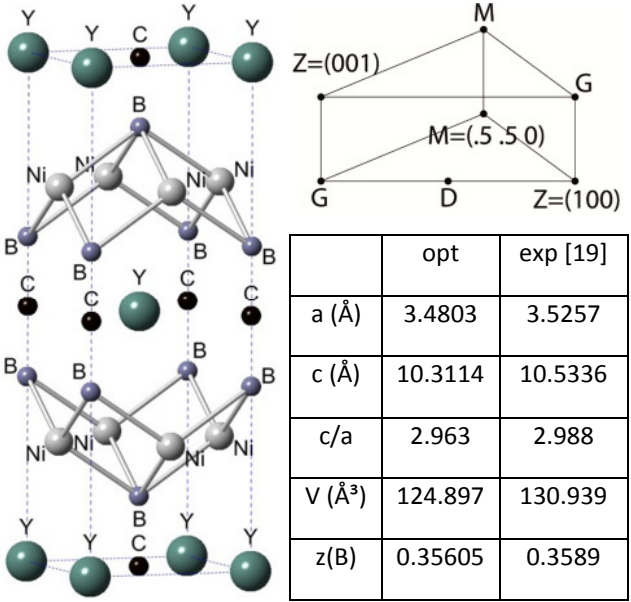

Figure 1
(a) Tetragonal unit cell of $YNi_2B_2C$ (I4/mmm [#139]) with color-coded atomic species: Y (green), Ni (grey), B (violet) and C (black). (b) Sketch of the Brillouin zone with high symmetry points. (c) Structural parameters of the optimized and experimental unit cell for $YNi_2B_2C$. The latter are taken from Ref. [19].

|        | opt     | exp [19] |
|--------|---------|----------|
| a (Å)  | 3.4803  | 3.5257   |
| c (Å)  | 10.3114 | 10.5336  |
| c/a    | 2.963   | 2.988    |
| V (Å³) | 124.897 | 130.939  |
| z(B)   | 0.35605 | 0.3589   |

monochromators to achieve high resolution. A fixed analyzer energy of 14.7 meV allowed us to use a graphite filter in the scattered beam to suppress higher orders. The wave vectors are given in reciprocal lattice units (rlu) of ($2\pi/a$ $2\pi/b$ $2\pi/c$), where a = b = 3.51 Å and c = 10.53 Å. Measurements were carried out both in the 100-001 and in the 110-001 scattering planes. The single crystal sample of $YNi_2B_2C$ weighing 2.26 g was mounted in a standard orange cryostat at LLB and in a closed-cycle refrigerator at FRM II, allowing us measurements down to T = 1.6 K and 3 K, respectively. However, all the results reported in this paper were performed at temperatures not lower than 20 K to avoid superconductivity-induced changes of phonon energies and line widths.

### IV. Results – Density Functional Theory

#### IV.1 Structural properties

$YNi_2B_2C$ crystallizes in a body-centered tetragonal structure (space group I4/mmm) shown in Fig. 1. The structure can be considered to be composed of $Ni_2B_2$ and YC sheets, which are alternately stacked along the c-axis. The C atom is covalently bound to its two neighboring B atoms, which leads to rather stiff bonds and a rigid B-C-B sub unit. The only internal degree of freedom of the structure is the z-coordinate of the B atoms. Selected high-symmetry directions of the Brillouin zone are sketched in Fig. 1. Due to the body-centering of the lattice, the point Z = (1, 0, 0) along the a*-axis is equivalent to the (0, 0, 1) point along the c*-axis [in units of ($2\pi/a, 2\pi/a, 2\pi/c$)].

The structural parameters of the fully optimized unit cell are given in the table included in Fig. 1. The lattice constants

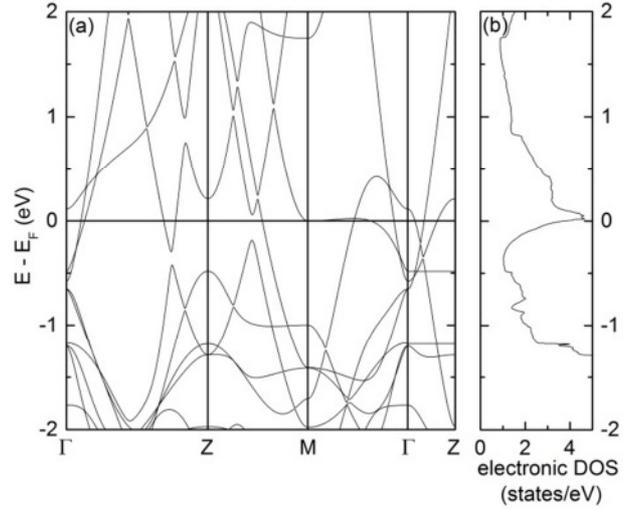

Figure 2
(a) Calculated electronic band structure of $YNi_2B_2C$ in the vicinity of the Fermi energy $E_F$. (b) Electronic density of states (eDOS) for $YNi_2B_2C$ near the Fermi energy $E_F$ (solid line).

are smaller than observed in experiment at room temperature [19] resulting in a 5% smaller volume of the unit cell. As we will discuss in section IV.3 the calculated lattice dynamical properties based on the optimized structure shows some deviations from experimental observations [20]. Therefore, we adopted the experimental lattice constants [19] as given in the table of Fig. 1 for our calculations. In the experimental part (section V.1) we will see that the calculated energies using room temperature lattice constants show good agreement even with phonon energies measured at T = 20 K.

The calculated superconducting transition temperature $T_c$ for the optimized structure is lower (5.8 K) than that for the experimental one (13.4 K). The reason for this large difference is probably related to the fact the c/a ratio for the optimized structure is about 1% lower than that for the experimental one, and it is known from experiment that $T_c$ depends very sensitively on this ratio [21].

#### IV.2 Electronic structure and Fermi surface

Details of the electronic structure are important for an understanding of the phonon anomalies discussed in Sec. IV.3. Although several band structure calculations have been carried out in the past [22,23], we briefly describe the outcome of our calculations here for the sake of a better readability. The band structure in the vicinity of the Fermi level is characterized by steep bands (Fig. 2a). A noticeable exception is the flat band along the Γ - M direction, which is mainly derived from Ni-3d states. The flat part of this band coincides almost perfectly with the Fermi energy, which raised speculations that this band plays a significant role for superconductivity [22].

The total electronic density of states (eDOS) is plotted in Fig. 2b. In $YNi_2B_2C$, the Fermi level is located in the left wing of a large peak, which results in a high $N(E_F)$ = 3.74 states/eV. This large value is mainly due to *3d*-derived Ni states. However, Y-, B- and C-derived states are present as well. The eDOS peaks 25 meV above $E_F$ with 4.6 states/eV. This peak is due to the very flat band along the Γ-M direction discussed above. Its position and height are in good agreement with previous reports [23].

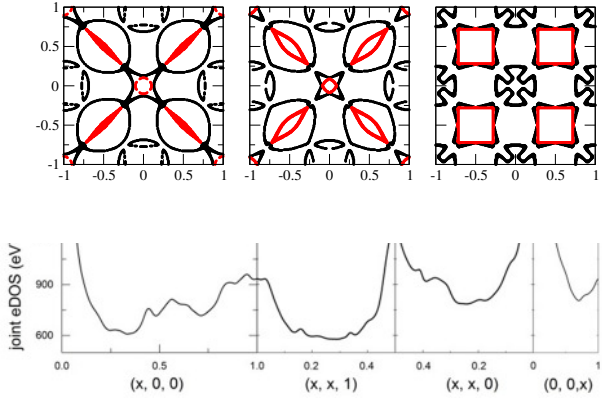

Figure 3
*(top)* Cuts through the Fermi surface (FS) parallel to the $k_x$-$k_y$ plane. Left, middle and right panels are for $k_z = 0$, 0.25 and 0.5, respectively (all k values given in r.l.u.). Black and red lines are parts of the FS derived from two bands crossing $E_F$ (see text). *(bottom)* Joint electronic density of states (eDOS) along various high symmetry direction of the Brillouin zone.

In the context of phonon anomalies discussed below, it is instructive to consider the Fermi surface topology in more detail. Cuts through the FS parallel to the a*-b* plane are plotted in Fig. 3 for three values of $k_z = 0$, 0.25 and 0.5 in an extended zone scheme. The FS is derived from three bands, #21–#23. The largest contributions come from band #21, which has Ni 3d character (see discussion on eDOS). A common feature related to this band is an ellipsoid-shaped FS centered at the M point for $k_z = 0$, which evolves for increasing $k_z$ into an almost square-like form. Furthermore, a small star/clover-shaped FS sheet centered at the (0, 0, z) direction and small pockets around (1, 0, z) are present. The second band (#22) (red parts in Fig. 3) gives also a square FS at $k_z = 0.5$, which shrinks rapidly for decreasing $k_z$. The third band (#23) produces small pockets around Γ (not shown in Fig. 3).

At $k_z = 0.5$, we see parallel sections of the Fermi surface connected by **q** = (0.56, 0, 0). However, this nesting is limited to $k_z$ values close to 0.5: The parallel sections disappear rapidly going away from $k_z = 0.5$. Calculating the wave vector dependence of the electronic phase space (joint electronic density of states) along various high-symmetry direction of the Brillouin zone, we found that the overall phase space is enhanced around **q** = (0.56, 0, 0), but this enhancement as well as the one at the Z point, i.e. **q** = (1, 0, 0) is weak. The strongest maximum apart from the Γ point is observed at the M point, i.e. **q** = (0.5, 0.5, 0) (Fig. 3). We will discuss how major phonon anomalies and strongly wave vector dependent EPC discussed below may be related with enhancements in the electronic phase space below.

### IV.3 Lattice dynamics

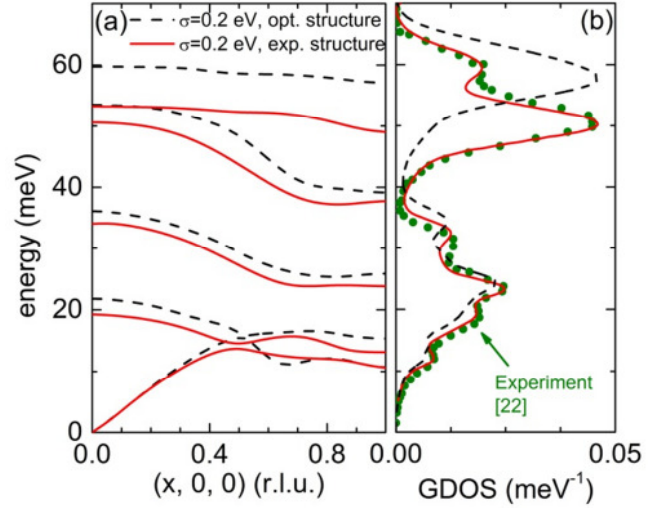

Figure 4
(a) Dispersion of c-axis polarized phonon modes along the (100) direction and (b) generalized phonon density of states (GDOS) using the optimized (dashed lines) and experimental structure (solid lines). Both calculations use a Gaussian smearing of σ = 0.2 eV (see text). The results for the GDOS are convoluted with an experimental energy resolution in order to be compared to data (dots) obtained via neutron scattering [22].

The structure of the vibrational spectrum for $YNi_2B_2C$ can be understood from its basic structural building blocks. To a first approximation, the unit cell can be considered to be built up from a B-C-B double dumbbell and three heavy atoms ($YNi_2$). The stiff bonds and light atom masses of the $B_2C$ double dumbbell give rise to two high-energy modes related to symmetric and antisymmetric bond-stretching vibrations of the $B_2C$ dumbbell along the c-axis. They have energies near 160 meV ($A_u$) and 100 meV ($A_{1g}$), respectively, which are well separated from the remaining vibrations lying below 70 meV. The lower energy part of the spectrum consists of one bond-bending and 6 external vibrations of the $B_2C$ unit (3 translations and 3 librations) against the heavy atoms in the energy range of approximately 40–70 meV. The remaining 9 vibrational modes are dominated by the heavy atoms and have energies below 40 meV.

Before we discuss calculated phonon dispersion curves in detail, we emphasize again that our calculations were done using the experimental lattice parameters [19] as opposed to a structure fully optimized within our theoretical framework. This choice was motivated by a comparison of the different calculated results with experimental observations available at the beginning of our investigation, i.e. the phonon density of states determined by inelastic neutron scattering as reported in [20]. Figure 4 compares phonon properties based on the fully optimized and experimental lattice parameters for the dispersions along the (100) direction. As expected we find generally higher energies for the optimized structure due to the smaller unit cell volume. The difference is most pronounced for the branch starting at 60 meV in the fully optimized calculation. Fig. 4b shows the comparison of the corresponding generalized phonon density of states with the experimental data extracted from [20]. The agreement with the latter is much better for the calculation using the experimental lattice parameters over the entire energy range and, in particular, for the pronounced peak located around 50 meV in the experimental data. This is also corroborated by our own experiments (see below) and previously published results for the high energy branches at E ≈ 100 meV and above ([11]). Therefore, here we present only results obtained for the experimental lattice parameters unless stated otherwise. Finally, we note that both the experimental lattice constants in Ref. [19] and the phonon DOS in Ref. [20] were reported for T = 300 K, and no experimental data were available at that time for lower temperatures. For this reason, we continued with the room temperature lattice constants. We note that the lattice constants of YNi$_2$B$_2$C decrease by less than 0.3 % on cooling from room temperature to a few degrees Kelvin [24].

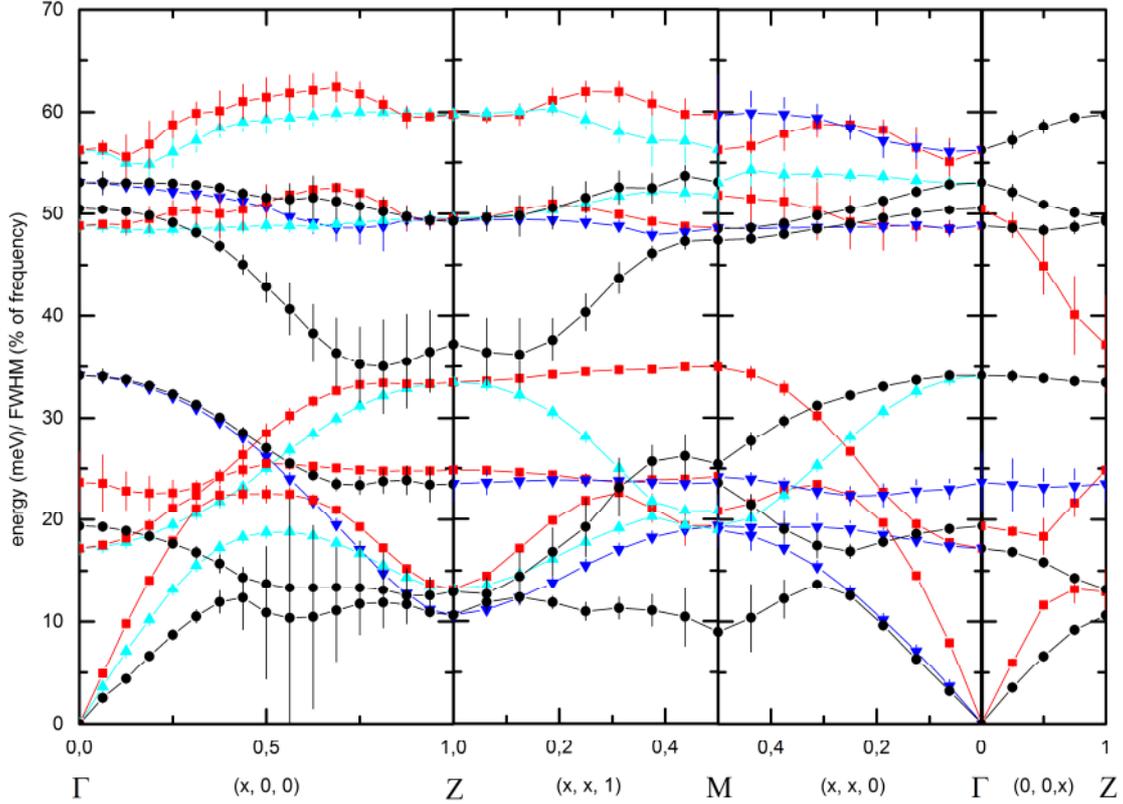

Figure 5
Calculated phonon dispersions and phonon line widths along the high-symmetry directions Γ – Z – M – Γ - Z for YNi$_2$B$_2$C. Symbols represent the phonon frequencies in meV, vertical bars show the value of the calculated electronic contribution to the phonon line width divided by the respective phonon frequency (the FWHM are given in % of the phonon energy, i.e. a vertical bar going from 0 to 20 [on the left hand scale] for a phonon at E = 10 meV denotes a FWHM of 2 meV). We note that each wave vector position was calculated individually in DFPT. Connecting lines are only guides to the eye. Different colors and symbols denote phonons of different symmetry classes. c-axis polarized phonons are shown in black dots except along the (001) direction, where they are shown as red squares. Note that the dispersion of the two highest branches at E ≈ 100 meV and 160 meV are subject of a different publication [11] and are not shown in this plot.

The complete phonon dispersion curves along various high-symmetry directions are displayed in Fig. 5. Only the branches below 70 meV are shown. The nearly flat branches near 100 meV and 160 meV have been discussed in detail previously [11]. Phonon properties were calculated in a dense three dimensional grid in momentum space and additional wave vector values were added for lines of particular interest, e.g., along the (100) direction. We emphasize that the results shown in Fig. 5 do not involve any interpolation in order to allow a clear assessment of the accuracy of our calculations for the reader. The calculated electronic contribution to the phonon line width (vertical bars), which will be discussed in the next subsection, is shown for each mode as well.

Inspection of Fig. 5 reveals a strong downward dispersion of the phonon mode starting at 52 meV at the zone center to 36 meV at the Z point. Furthermore, two pronounced anomalies appear in the transverse acoustic (TA) branches along the (100) and (110) directions. All three modes are mainly c-axis polarized and show a strong coupling to the electronic system. Typically such investigations run into the question of whether phonon softening and/or broadening originate from electron-phonon coupling to conduction electrons or to anharmonicity. In $YNi_2B_2C$ anharmonicity can be ruled out with certainty for the acoustic branches, because of the previously reported effect of the superconducting gap on the phonon lineshapes below $T_c$. This highly nontrivial effect can be very well modeled by theory based exclusively on electron-phonon origin of the normal-state broadening [9].

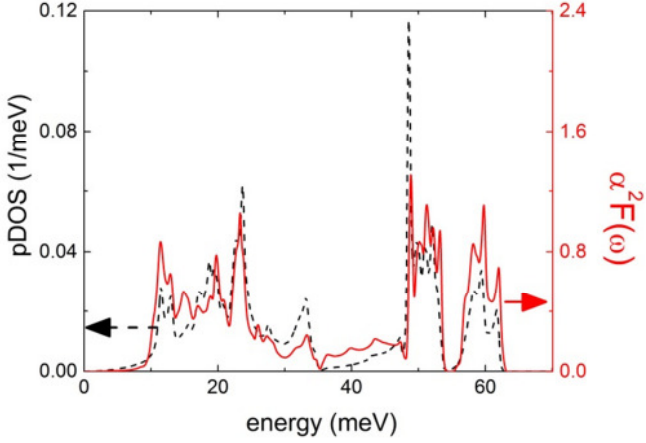

Figure 6
Calculated phonon density of states (PDOS, dashed line) and isotropic Eliashberg function $\alpha^2F(\omega)$ (solid line) of $YNi_2B_2C$.

### IV.4 Phonon-mediated superconductivity

In this section, we discuss details of the electron-phonon interaction. The first theoretical investigation of EPC and the phonon line widths in a superconductor was reported in the late 1970s for elemental Niobium [5], a much simpler system with respect to its atomic as well as electronic structure. In particular, the phonon eigenvectors were not yet considered realistically. We will demonstrate that in $YNi_2B_2C$ the presence of EPC and concomitant large phonon line widths are closely connected with particular phonon eigenvectors.

The phonon-mediated pairing interaction is described by the isotropic Eliashberg function [25]

$$\alpha^2F(\omega) = \frac{1}{2\pi N(0)} \sum_{q,\lambda} \frac{\gamma_{q,\lambda}}{\omega_{q,\lambda}} \delta(\omega - \omega_{q,\lambda}), \quad (1)$$

where $\omega_{q,\lambda}$ denotes the energy of the phonon mode $(q,\lambda)$, $N(0)$ is the electronic density of states (per atom and spin) at the Fermi energy, and $\gamma_{q,\lambda}$ is the electronic contribution to the phonon line width

$$\gamma_{q,\lambda} = 2\pi \omega_{q,\lambda} \sum_{k,\nu,\nu'} \left| g_{k+q\nu',k\nu}^{q,\lambda} \right|^2 \delta(\epsilon_{k\nu} - \epsilon_F) \delta(\epsilon_{q+k\nu'} - \epsilon_F) \quad (2).$$

Here, $g$ denotes the screened EPC matrix element. Within the perturbational approach to the lattice dynamics, $g$ is directly accessible from quantities obtained in the calculation of the dynamical matrix.

Fig. 5 includes detailed results for the energy and momentum dependence of $\gamma_{q,\lambda}$ for the various phonon branches. The calculated values of the FWHM $2\gamma_{q,\lambda}$ divided by the respective phonon energy are shown as vertical bars. We have chosen this way of presentation as these values are proportional to the contribution of the different modes to the Eliashberg function $\alpha^2F(\omega)$ (see Eq. 1). The latter is plotted in Fig. 6 together with the phonon density of states (PDOS). The two quantities are plotted such that the graphical integral is the same for both, i.e. the area under the curves in the figure is the same. It is clear that for some particular energies, e.g., E = 10 - 15 meV and around 40 meV, the phonons have a relatively large contribution to $\alpha^2F(\omega)$ compared to their spectral weight in the PDOS. This indicates a strong EPC of these modes as $\alpha^2F(\omega)$ can be approximately seen as a PDOS weighted by the energy dependent EPC strength of all modes.

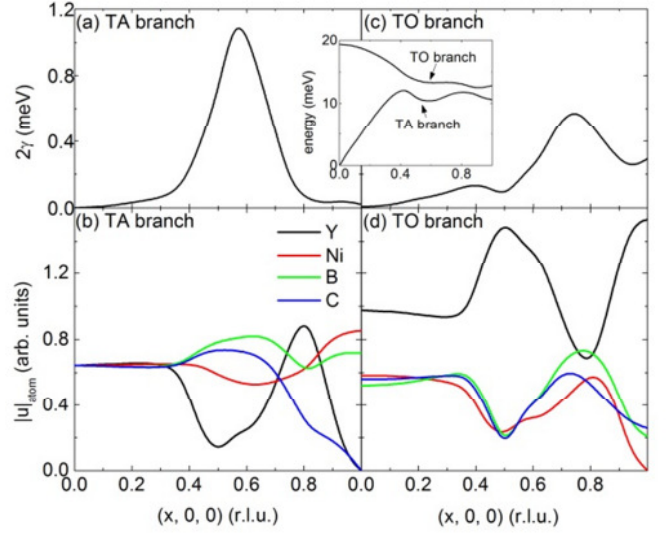

Figure 7
Calculated electronic contribution to the phonon line width $2\gamma$ (upper panels) and moduli of atomic amplitudes (lower panels) for the c-axis polarized TA and first TO phonon branches along the crystallographic (100) direction. The inset shows the calculated phonon dispersion (see also Fig. 5).

A particular result of our work has been discussed already in detail in a previous publication [11], namely that the branches below 70 meV contribute the most to the total EPC coupling strength. In fact, less than 3% of the total EPC constant $\lambda = 2 \int_0^\infty d\omega \frac{\alpha^2F(\omega)}{\omega}$ stem from the branches above 70 meV, but two thirds of $\lambda$ originate from modes below 30 meV. At the same time, our calculations predict that two thirds of the total EPC are due to contributions from the light atoms B and C. The latter two statements, i.e., strong coupling in low energy branches and a major contribution to EPC from the light atoms, seem to be somewhat contradictory as low energy vibrations are usually dominated by the heavy atoms. Yet, they are not really contradictory because our calculations revealed that the EPC not only renormalizes the phonon energies, but also influences the phonon eigenvectors, giving rise to relatively large vibrational amplitudes of B and C atoms in the low energy branches. For a more detailed discussion, we will focus on the c-axis polarized branches along the (100) and (110) directions, which have large values of $2\gamma$ around wave vector $\mathbf{q}$ = (0.56, 0, 0) and (0.5, 0.5, 0) (Fig. 5).

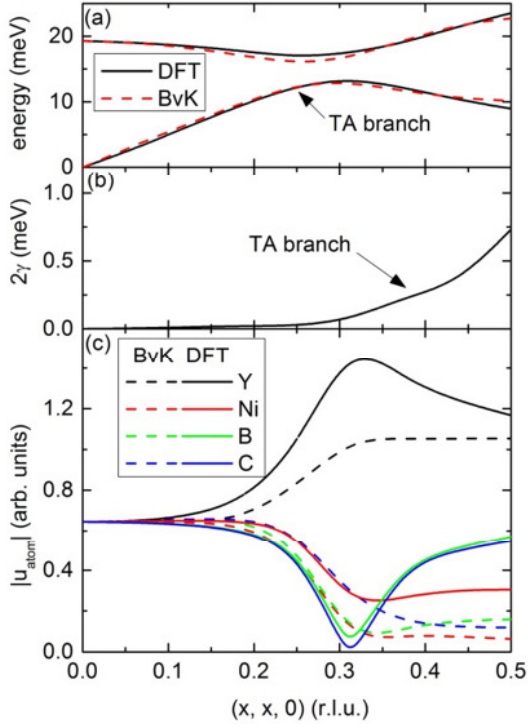

Figure 8
Calculated dispersion, electron-phonon coupling related line widths and atomic amplitudes for the c-axis polarized TA branch along the crystallographic (110) direction. *(a)* Calculated phonon dispersion using DFT (solid lines) and a Born-von-Karman model (BvK, dashed lines, see text). *(b)* Within DFT calculated electronic contribution to the phonon line width $2\gamma$ (FWHM) for the TA branch. *(c)* Amplitudes of the different atomic species for the TA branch as function of wave vector. Again, we plot results from DFT (solid lines) and BvK calculations (dashed lines).

Fig. 7 depicts detailed results of $2\gamma$ and the phonon displacements pattern on the TA and first transverse optical (TO) branches along the (100) direction, which show both a strong wave vector dependence of $2\gamma$ (Figs. 7a,c). The inset shows the dispersion of the two branches with clear indications for an exchange of eigenvectors near $\mathbf{q} = (0.4, 0, 0)$. This exchange is reflected in the calculated displacement patterns (Figs. 7b,d). At x = 0.4 the Y amplitude of the TA (TO) mode is strongly reduced (increased), whereas B and C amplitudes show the opposite behavior. Going further towards the zone boundary, we see a second exchange around x = 0.8 with a strong peak (dip) in the Y amplitude of the TA (TO) branch. In both the TA and TO branches we see that strong EPC is present when the B and C movements are large (Figs. 7a,b). This seems not to be the case in the TA branch near the zone boundary, where EPC becomes very small, but the B amplitude stays large. However, the TA mode acquires a strong B motion along the a-axis for x ≥ 0.8, whereas the movement along the c-axis goes to zero at x = 1 (similar to C in Fig. 7b). Strong EPC in the TA branch is, therefore, clearly connected to the concomitant presence of increased c-axis polarized B and C amplitudes and relatively weak Y displacements. In the wave vector region x ≥ 0.4, we see the same behavior for the first TO branch, i.e. peaks in the amplitudes of B and C together with a pronounced minimum of the Y amplitude at the same wave vector value where $2\gamma$ shows its maximum (Figs. 7c,d).

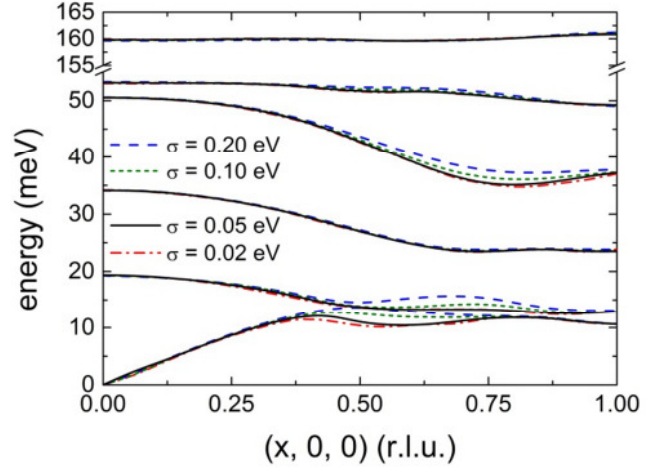

Figure 9
Dispersion of c-axis polarized phonon modes along the (100) direction using four different values for the smearing σ = 0.02 eV (dash-dot), 0.05 eV (solid), 0.10 eV (short-dash) and 0.2 eV (long-dash).

Focusing now on the TA branch along the (110) direction, we present a comparison between DFPT and a Born-von-Karman (BvK) model. The latter was fitted to the dispersion obtained via DFPT in order to find out how well a more simplistic model (without EPC) can describe not only the dispersion but also the displacement pattern. We obtained a good match between the two different sets of phonon dispersions (Fig. 8a). Yet, we find a qualitative difference in the atomic displacements at $\mathbf{q} = (x, x, 0)$ for x > 0.3. Whereas the BvK model predicts continuously low B and C amplitudes at these wave vectors, the DFPT shows a substantial increase for $0.3 \leq x \leq 0.5$. As we will show in the experimental section, the different displacement patterns lead to very different structure factors for the TA phonon at x = 0.5, which we were able to check experimentally in multiple Brillouin zones. The result agrees very well with the DFPT calculation and contradicts the BvK model (Tab. 2). Hence, our analysis demonstrates that DFPT is able to predict the lattice dynamics in YNi$_2$B$_2$C in great detail. On the other hand, we know from our previous discussion that large c-axis polarized B and C amplitudes are most important for a phonon mode in order to exhibit strong EPC. Comparing the atomic amplitudes with the calculated wave vector dependent $2\gamma$ we see that the latter starts to increase right where we have the turnaround of the B and C amplitudes (Figs. 8b,c). In summary, we conclude that DFPT predicts strong EPC in various c-axis polarized phonon branches. The EPC is coupled to unusually large amplitudes of the B and C atoms. Such amplitudes can be expected for the TO branch starting out near 50 meV at the zone boundary and dispersing down to 36 meV at the Z point (see [11]). However, it is a non-trivial result for the discussed acoustic modes and is confirmed by experiment (see section V.3).

| σ (eV) | $\omega_{log}$ (meV) | N(0) (1/eV) | λ |
|---|---|---|---|
| 0.02 | 23.415 | 2.07051 | 0.79878 |
| 0.05 | 23.171 | 1.92055 | 0.81085 |
| 0.1 | 24.363 | 1.72543 | 0.74707 |
| 0.2 | 25.866 | 1.47915 | 0.63303 |

Table 1
Parameters relevant for superconducting properties calculated in DFPT for various smearing constants σ = 0.02 eV, 0.05 eV, 0.1 eV and 0.2 eV. Reported are the effective frequency $\omega_{log}$, the electronic DOS at the Fermi energy per spin and cell N(0) and the EPC constant λ.

All calculated results, which we have presented so far, used a numerical smearing constant σ = 0.05 eV. DFPT calculations require such a numerical smearing σ in **k** space due to the finite momentum mesh used. This smearing of the electronic states resembles that caused by finite temperatures, and thus a variation of σ can be used to study temperature effects.

It was already noted in [26-28] that the temperature dependence of Kohn anomalies can be simulated with different values of σ. We used the same procedure to study the temperature dependence of the lattice dynamical properties of YNi$_2$B$_2$C. The dispersion along the (100) direction calculated with σ = 0.02 eV, 0.05 eV, 0.1 eV and 0.2 eV is shown in Fig. 9. For the sake of clarity, we restrict the plot to the c-axis polarized branches. Note that the results for σ = 0.05 eV are also shown in the left panel of Fig. 5 (black lines) including the EPC strength as vertical bars. Hence, we see that only phonon modes having strong EPC are sensitive to a variation in σ. For instance, the anomaly in the TA mode discussed above is more pronounced for smaller values of σ. The same trend is also observed for the other modes in Fig. 9. Interpreting σ as a temperature it follows that effects due to EPC should be more pronounced at lower temperatures which we find to be in good agreement with experiment. However, σ ≡ 2.12*$k_B$T, would mean that very high temperatures are necessary to bring about sizeable changes of phonon energies and line widths. Yet in experiment similar changes are observed in a temperature range which is one order of magnitude smaller (see sect. V.2 and Fig. 14). It is not obvious if the observed temperature dependence of the phonons is related to the calculated one. An important piece of evidence in favor of them being related is that the q-region where the calculated and experimental temperature-dependences are the strongest are nearly the same. One possible reason for the lower temperatures at which phonon softening occurs in the experiments than in the calculations that the calculations ignore the thermal disorder of the lattice, which can have profound effects on the phonon excitations [29]. Elucidating these issues is an important direction for future research but is beyond the scope of this paper.

The result of increased EPC in calculations with smaller values of σ can be summarized by two particular values, i.e. the effective phonon energy [30]

$$\omega_{log} = exp\left(\frac{2}{\lambda}\int_0^\infty d\omega \frac{\alpha^2 F(\omega)}{\omega} ln(\omega)\right) \quad (3)$$

and the overall EPC constant [30]

$$\lambda = 2\int_0^\infty d\omega \frac{\alpha^2 F(\omega)}{\omega} \quad (4)$$

Calculated numbers of these quantities using different values of σ are listed in Tab. 1 along with N(0), the electronic density-of-states per spin and unit cell at the Fermi energy. Going from σ = 0.2 eV to 0.05eV we see an increase of λ, i.e. a strengthened EPC. For σ = 0.02 eV, λ is again slightly reduced and also the trend in the effective energy to become smaller for reduced σ is reversed. As we will see in section V.2 the latter is not reflected in the single phonon energies of some particularly strong coupling phonons. We surmise that our **k**-point mesh was too coarse to achieve a reliable convergence of our calculations when using a smearing parameter as small as σ = 0.02eV. Therefore, we use generally the results for σ = 0.05eV throughout this report, for which we are confident about the numerical convergence.

Based on the results of DFPT listed in Table 1, we can calculate the superconducting transition temperature $T_c$ of YNi$_2$B$_2$C by solving the gap equation. In order to do this, however, we have to assume a value for μ*, the effective electron-electron interaction, which is not given by our calculations. Using a typical value μ* = 0.13, we find $T_c$ = 13.4 K (for σ = 0.05 eV). We note that a calculation based on the optimized crystal structure (see table in Fig. 1) yields a transition temperature of only 5.8 K using the same μ*. This emphasizes that the physical properties of YNi$_2$B$_2$C are much better described by our DFPT calculations when employing the experimental lattice constants.

### V. Results - Experiment

### V.1 Phonon dispersions

Although superconductivity in YNi$_2$B$_2$C was discovered nearly two decades ago, no systematic investigation of the phonon dispersion curves were undertaken so far. As mentioned in the introduction, phonon studies of YNi$_2$B$_2$C focused on the dramatic superconductivity-induced changes of the line shape of the TA mode at **q** = (0.5, 0, 0) [4,31] and, hence, only the dispersion curves of this TA and the first TO modes were reported for the normal state [32,33].

The phonon energies in YNi$_2$B$_2$C span a wide range exceeding 100 meV [11,20]. Our inelastic neutron scattering experiments on triple axis spectrometers using thermal neutrons were restricted to the energy range up to 70 meV, which contains 16 out of a total of 18 phonon branches (Fig. 5). A detailed study of the branches with E ≥ 100 meV using neutron time-of-flight spectroscopy is published elsewhere [11]. In the (100) direction all phonon branches accessible in the 100-001 scattering plane were measured. The same is true for the branches in the (110) direction, which were measured in the 110-001 scattering plane. There was, however, the problem that the structure factor of the TA (110) phonon branch near the zone boundary, the so-called M point, is very low for all reciprocal lattice points τ contained in the 110-001 scattering plane. Fortunately enough, the endpoint of the TA (110) phonon branch could be investigated with a very high structure factor at **Q** = (0.5, 0.5, 7), i.e. a point contained in the 110-001 scattering plane, which is connected by a wave vector (-0.5, 0.5, 0) with reciprocal lattice points above and below the scattering plane, i.e. (1, 0, 7) and (-1, 0, 7). Neighboring points in **q** were accessible by tilting the sample by a few degrees.

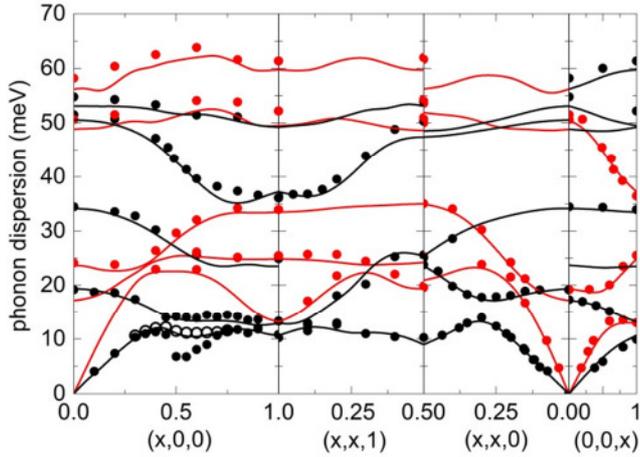

Figure 10
Calculated (lines) and observed (filled dots) phonon dispersion in YNi$_2$B$_2$C at T = 20K along various crystallographic high symmetry directions. Open circles shown for the TA mode along the (100) direction were measured at T = 250 K. Branches shown in red or black refer to phonons of predominantly longitudinal or transverse polarization, respectively (same colour code as in Fig. 5).

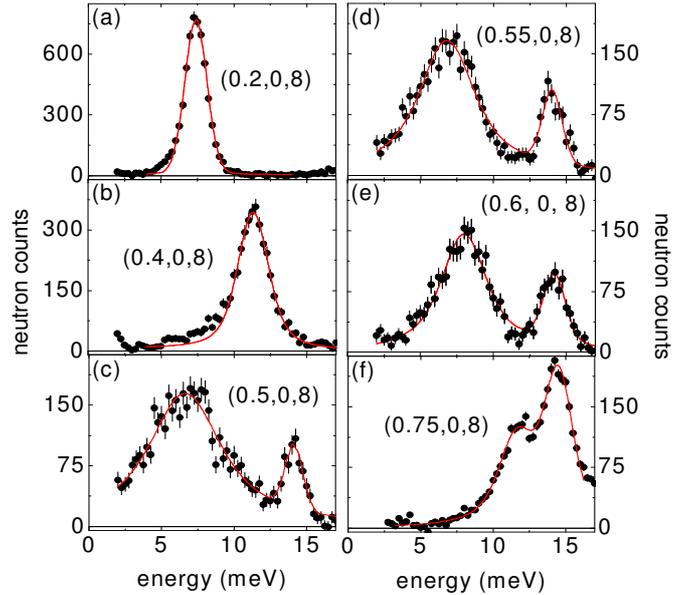

Figure 11
Background subtracted energy scans taken at **Q** = (0.2 - 0.75, 0, 8) and T = 20K. Lines are fits using Lorentzians convoluted with the experimental resolution.

A summary of the experimentally observed phonon energies is given in Fig. 10. Lines represent the DFPT calculations as presented in Fig. 5. Branches shown in red or black refer to phonons of predominantly longitudinal or transverse polarization, respectively. As explained in section IV, we did not use the optimized crystal structure, but rather employed the experimental lattice constants of YNi$_2$B$_2$C reported at room temperature. This produced a very good agreement with PDOS data (Fig. 4b, [20]) taken at room temperature as well.

Inspection of Fig. 10 shows that there is in general a very good agreement between theory and experiment. It is true that the calculated energies for ab-plane polarized phonon modes are slightly too low, but the difference does not exceed 3% of the predicted values. Using low temperature lattice constants instead of room temperature ones would have somewhat improved the situation. Since the c-axis lattice constant was reported to be practically temperature independent [24], using low temperature lattice constants would not have spoiled the very good agreement over the whole energy range for c-axis polarized phonons (results for E > 70 meV are published in [11]).

However, a clear difference between theory and experiment is observed for the anomaly in the TA branch at 20K, which is much more pronounced than predicted in agreement with previous work [4,33,34], although the room temperature data for the TA 100 branch fit well to the reported ones at T = 150 K and 620 K [33]. All other modes predicted to have strong EPC show a rather good agreement with the calculated energies. Fig. 2 includes data for the TA branch in the (100) direction taken at T = 250 K to demonstrate that these energies (open circles in Fig. 10) are in much better agreement with theory than the low temperature data for this particular branch. The temperature dependence of energy and line width of this particular mode is extremely strong and will be discussed in more detail below.

### V.2 Electron-phonon coupling

A large part of our beam time was devoted to the determination of the intrinsic line width of phonons, for which DFPT predicted strong EPC. In the following we label the intrinsic phonon line width obtained experimentally as Γ (HWHM) in order to avoid confusion with the calculated electronic contribution to the phonon line width γ in DFPT.

We note that a determination of the EPC related line width is quite a challenge for inelastic neutron scattering because they rarely exceed a few % of the phonon energy, and hence are comparable to or even smaller than the experimental resolution. Therefore, we have chosen high-resolution experimental set-ups and paid particular attention to achieve a good knowledge of the experimental background. The observed line widths had to be corrected for resolution effects. The experimental resolution was calculated using a standard formalism and the values were checked by measurements of phonons having a very low EPC strength.

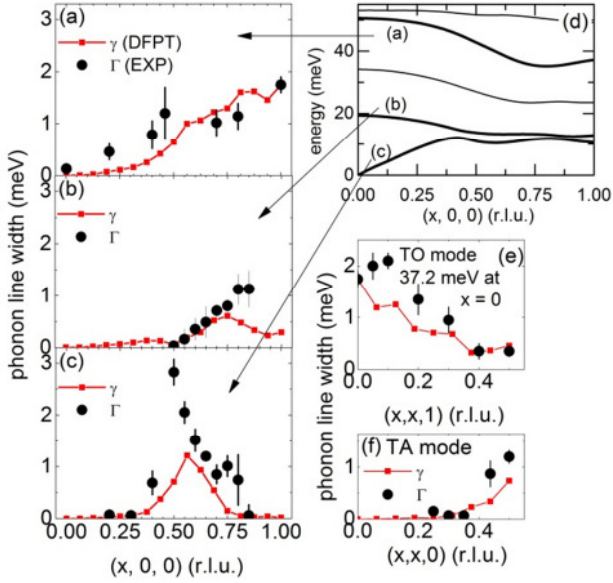
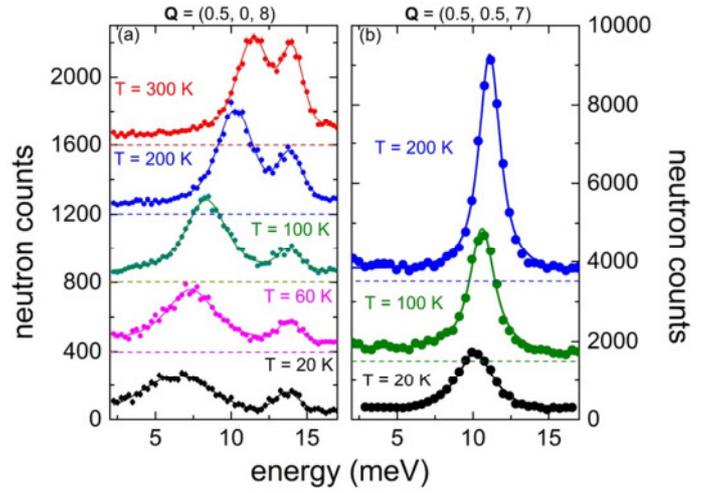

Figure 12
Observed (dots) and calculated (squares) wave vector dependent phonon linewidths (HWHM) for c-axis polarized phonon modes measured at wave vectors **q** = (x, 0, 0), (x, x, 1) and (x, x, 0) shown in panels (a)-(c), (e) and (f), respectively. Panel (d) shows the dispersion of c-axis polarized modes along the (100) direction. Thick lines denote the branches for which the calculated and observed line widths are shown in panels (a)-(c).

We systematically measured all accessible branches, for which theory predicted a detectable line width. There were only a few examples where the expected large line width was not observed. For example, the zone center phonon at 24 meV, for which DFPT predicted a HWHM of 0.7 meV, showed no detectable intrinsic line width, i.e. has to be smaller than 0.3 meV.

In the following, we focus on three c-axis polarized phonon branches, for which DFPT predicted relative line widths of up to 25 % of the phonon energy (Fig. 5). We remind the reader that the importance of a single phonon for conventional superconductivity is given by the relative line widths, i.e. the ratio of the line width and energy, $\gamma/\omega$, in the Eliashberg equation (Eq. 1) [35]. For the sake of clarity, we present the results on EPC and the phonon eigenvectors for each of the three modes in special subsections.

**TA phonon branch in the (100) direction**

Energy scans showing peaks associated with the c-axis polarized TA phonon branch along the (100) direction are displayed in Fig. 11. The additional peaks appearing at the upper end of the energy range come from the first TO phonon branch. The TO peaks do not show up in panels a) and b) because their energies is beyond E=17 meV and the corresponding intensities are much weaker than for the TA phonon (for the TO dispersion see Fig. 10). Inspection of the scans reveals a sudden drop of the TA phonon energy when going from x = 0.4 to 0.5. At the same time, the phonon peak broadens massively. When x increases beyond 0.5, the

Figure 13
Energy scans taken at (a) **Q** = (0.5, 0, 8) and (b) **Q** = (0.5, 0.5, 7) (M-point) at various temperatures. Solid lines are fits using a Lorentzian convoluted with the experimental resolution. Dashed lines denote the base line for the respective scan.

TA phonons harden quickly and their line widths shrink considerably. The results for the wave vector dependent phonon line widths are summarized in Fig. 12c. These features are the attributes of a pronounced phonon anomaly which was already reported some time ago [4]. Our DFPT calculations reproduce the observed anomaly qualitatively but underestimate its strength in the reduction of the phonon energy (Fig. 10) as well as the phonon broadening (Figs. 12). Besides, the maximum broadening is calculated at x = 0.56 whereas observed at x = 0.5. The only moderate agreement between theory and experiment with respect to this phonon anomaly is a clear shortcoming of our DFPT calculations, but one has to bear in mind that this anomaly is apparently extremely sensitive to details of the electronic structure: on raising the temperature to T = 250 K, the dip in the TA phonon dispersion curve is strongly reduced and matches the calculated one (open circles in Fig. 10). In our calculations the maximum line width is predicted for the same wave vector as a strong Fermi surface nesting, i.e. at **q** = (0.56, 0, 0). The experimental results, however, imply that the nesting vector is **q** = (0.5, 0, 0).

The temperature dependence of the phonon anomaly is documented in more detail in Fig. 13. Apparently, the TA phonon energy at x = 0.5 increases substantially on raising the temperature. We note that for T > 200 K the energy increase is even somewhat masked by a mixing of eigenvectors with the next higher TO phonon mode. The energy increase is accompanied by a strong reduction of the phonon line width. We believe that these temperature effects are due to a smearing of the Fermi surface and a concomitant reduction of the EPC strength. This view is supported by DFPT calculations using different values of the smearing parameter σ (see section IV.4). As discussed in section IV.4, a variation of σ can be used to study temperature effects in a qualitative way. Indeed, a variation of σ leads to the same trends as in experiment as a function of temperature (Fig. 14 a,d). We remind the reader that the absolute values of σ are about one order of magnitude too large compared to experiment. One reason for the difference might be the lack of thermal motion of the ions in DFPT, which can be important for lattice dynamics at room temperature.

When going from x = 0.5 towards the zone boundary, the line width of the TA phonons decreases but that of the next higher TO phonons increases (Fig. 12 b,c). This phenomenon is also captured by our DFPT calculations, although again not quantitatively. We think that these variations of the line width are linked to an exchange of eigenvectors. This issue will be discussed in more detail in Section V.3.

**TO branches in the (100), (001) and (xx1) directions**

We note that parts of the data presented below (Fig. 12a,e) were already included in a previous report [11]. However, previously unpublished data warrant a presentation of results on the TO mode on its own, where we show the already published data for the sake of completeness.

In high energy branches, the predicted relative line width is largest for a c-axis polarized phonon with an energy of 36.1 meV at **q** = (1, 0, 0) (Fig. 5). From zone center to zone boundary, the corresponding phonon branch shows a strong downward dispersion starting as high as 51.5 meV at the zone center. We followed this branch also along the zone boundary line (xx1) (0 ≤ x ≤ 0.5), and found a maximum energy of 50.3 meV at (0.5, 0.5, 1) (Fig.10). Very good agreement between experiment and theory was found along all lines in **q** space studies in our experiment. Only the weak dispersion minimum predicted at **q** = (0.8, 0, 0) was not observed. Rather, the minimum energy was found at the zone boundary.

Our results for the phonon line width in the (100) and the (xx1) directions are summarized in Fig. 12a and e, respectively. Attempts to determine the phonon line width of the corresponding phonon branch in the (001) direction were unsuccessful because the momentum resolution of the spectrometer was insufficient for a branch having an extremely steep dispersion[2]. We observed a monotonic increase of the line width along the (100) direction from the zone center to the zone boundary. Along the (xx1) direction our data indicate a maximum of the line width around **q** = (0.1, 0.1, 1) followed by a strong decrease for larger wave vectors. However, in good agreement with DFPT results, the phonon peak was still not resolution-limited at q = (0.5, 0.5, 1).

The temperature dependence of the phonon energy and line width at **q** = (1, 0, 0) for 20 K ≤ T ≤ 250 K is shown in Fig. 14c. Again, we see a hardening of the phonon energy with increasing temperature but there was no simultaneous decrease of the phonon line width within the experimental error bars.

We evaluated DFPT calculations with different smearing constants σ also for this TO mode (see Fig. 14f). The calculations reproduce the trend in the phonon energy but underestimate the hardening in absolute numbers. DFPT predicts a slight dependence of the line width for σ in the range of 50 meV to 200 meV. This is not reflected in the experimental results, but the error bars of the intrinsic line widths are relatively large, too, with about ±0.5 meV and more, for which reason no definite conclusion can be drawn. Error bars were, however, small enough to exclude the strong increase of the linewidth at very low temperatures predicted by DFPT using σ = 20 meV DFPT. It might well be that the calculated result is unreliable because of convergence problems for such a small σ. (see also section IV.4).

---
[2] Note that $|c^*| \approx 1/3 \times |a^*|$.

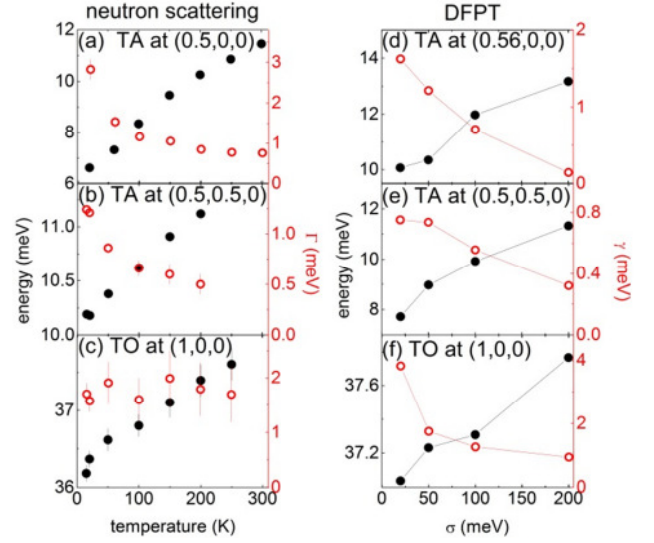

Figure 14
*Left column:* Experimentally observed temperature dependence of the phonon energy (filled symbols) and line widths (open symbols) of c-axis polarized modes at (a) **q** = (0.5, 0, 0), (b) **q** = (0.5, 0.5, 0) and (c) **q** = (1, 0, 0), respectively.
*Right column:* Calculated phonon energies (filled symbols) and line widths (open symbols) of c-axis polarized modes for different smearing constants σ = 20 meV, 50 meV, 100 meV and 200 meV at (d) **q** = (0.56, 0, 0), (e) **q** = (0.5, 0.5, 0) and (f) **q** = (1, 0, 0), respectively.

**TA branch in the (110) direction**

A previously unknown phonon anomaly was predicted for the TA phonon branch at the zone boundary in the (110) direction, the so-called M-point. The dispersion of this c-axis polarized low energy phonon branch at T = 20 K has been studied in the Brillouin zone around **τ** = (0, 0, 8) for **q** ≤ (0.35, 0.35, 0). The experimental energies are in very good agreement with theory (Fig. 10). Since there is an anti-crossing with the first TO branch around **q** = (0.3, 0.3, 0) transferring nearly all spectral weight to the upper branch, the dispersion and the line width of the TA branch for **q** > (0.35, 0.35, 0) had to be studied in a different Brillouin zone, i.e. around **τ** = (1, 0, 7). This choice of **τ** was suggested by the DFPT results and indeed, the structure factors were very favorable for the TA phonons near the zone boundary, i.e. **Q** = (0.5, 0.5, 7).

The experimental and theoretical results for the low temperature line width, T = 20 K, of the TA phonon branch are displayed in Fig. 12f. There is qualitative agreement between the calculated and the observed behavior, but similarly to the case of the TA phonon mode along the (100) direction, the absolute value of the maximum line width is underestimated by DFPT by a factor of 1.6. We remind the reader that DFPT showed that the presence of B and C amplitudes in this zone boundary mode is important for the large line width due to EPC (Fig. 8). The good agreement between the observed and predicted wave vector dependences gives further credence to this result.

Raw data for three different temperatures for the zone boundary TA phonon are shown in Fig. 13b. Obviously, the phonon's energy and line width change significantly in the temperature range from T = 20 K to 200 K. A summary of the results obtained at various temperatures is given in Fig. 14b. The behavior is qualitatively similar to that of the TA phonon at **q** = (0.5, 0, 0) (Fig. 14a), however the absolute effects are much smaller. A less pronounced temperature dependence goes along with the fact that we find no clear nesting for the wave vector **q** = (0.5, 0.5, 0) (see section IV.3).

Comparing the experimental results with calculations using different smearing constants σ for this particular TA phonon again gives the right trends. However, this time DFPT overestimates the observed temperature effect on the phonon energy within the investigated range of smearing constants, i.e. 3 meV in DFPT but only 1 meV in experiment. On the other side, the line width increases more strongly in experiment than calculated.

### V.3 Phonon eigenvectors

The experimental investigation of phonon eigenvectors is an ambitious goal and is therefore rarely done. Ideally, it requires studying the phonon of interest in a large number of Brillouin zones. This means that Brillouin zones have to be included for which the structure factors of the particular phonon are quite small. Therefore, a precise knowledge of the experimental background becomes crucial. In addition, problems might arise from an accidental degeneracy or near-degeneracy of other phonons. Consequently, the number of symmetry-inequivalent Brillouin zones for such a study is quite limited and hence phonon eigenvectors can be determined by experiment only in a very approximate way.

Our interest in the atomic displacement patterns of the TA phonons at **q** = (0.5, 0, 0) and (0.5, 0.5, 0) originated in their strongly anomalous behavior but also in the very unusual eigenvectors predicted by DFPT (see section IV.4, Figs. 7 & 8).

### TA and 1st TO branches in the (100) direction

In the long wavelength limit, the eigenvectors of the acoustic phonons are determined by symmetry. In the case of the TA (100) branch exhibiting the pronounced phonon anomaly, all atoms move in-phase along c. DFPT predicts that this character is retained up to x ≈ 0.4 from whereon the acoustic character is transferred to the first TO branch (see Fig. 7). Measurements of the TA and TO phonons at **q** = (x, 0, 8) largely confirm this scenario with large opposite changes in the intensities of the TA and first TO branches (Fig. 15a,b). DFPT predicts a further exchange of eigenvectors between the TA and TO branches on approaching the zone boundary, but this prediction could not be confirmed with certainty. We note that the two branches become very close in energy near (1,0,8) which makes it difficult to separate the peaks associated with the two branches.

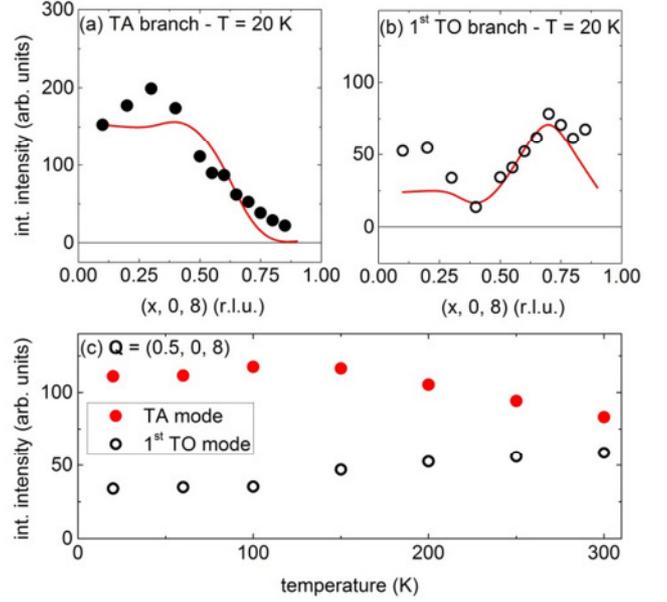

Figure 15
*(a)(b)* Symbols denote integrated intensities of the TA and first TO phonon branches measured in the Brillouin zone adjacent to **τ** = (0, 0, 8) at T = 20 K. Intensities were corrected for the Bose factor and the factor 1/ω in the phonon cross section. Lines are DFPT structure factor calculations scaled to the intensity of the TA phonon at x = 0.1. (c) Integrated intensities of TA and first TO modes at **Q** = (0.5, 0, 8) as function of temperature. Data were corrected for the Bose factor and 1/ℏω in the scattering cross section.

Temperature dependent measurements of phonon intensities at **q** = (0.5, 0, 8) lend support to the scenario of an exchange of eigenvectors between the TA and TO branches. On raising the temperature from 10 K to 300 K, we observe a significant decrease of the TA phonon intensity and a concomitant gain of the TO one (Fig. 15c). As was discussed above, the TA phonon frequency increases strongly on heating, bringing it closer and closer to the TO frequency, which in turn gives rise to a stronger hybridization.

A strong hybridization between the TA and first TO branch along the (100) direction was already discussed by Stassis and co-workers, in particular, for LuNi$_2$B$_2$C [36]. We agree that the hybridization of the two branches in YNi$_2$B$_2$C is temperature dependent. However, as was discussed earlier by us [9], the hybridization is not the origin for the anomalous line shape of the TA mode in the superconducting state.

### TA phonon at q = (0.5, 0.5, 0)

| Q | TA | | |
|---|---|---|---|
| | Exp | DFPT | BvK |
| (0.2, 0.2, 4) | 35 | 35 | 35 |
| (0.5, 0.5, 5) | 7 | 6 | 5 |
| (0.5, 0.5, 6) | 1 | 0.1 | 10 |
| (0.5, 0.5, 7) | 50 | 52 | 16 |
| (0.5, 0.5, 8) | <1 | 0.01 | 9 |

Table 2
Measured (EXP) and calculated (DFPT, BvK model) structure factors of the transverse acoustic (TA) phonon at **q** = (0.5, 0.5, 0) in different Brillouin zones at T = 20 K. Calculated structure factors were normalized to the experimental value at **Q** = (0.2, 0.2, 4).

Table 2 summarizes the integrated intensities of the TA phonon at the M-point, i.e. **q** = (0.5, 0.5, 0) observed at low temperatures in different Brillouin zones and compares them to the calculated structure factors. The experimental values are corrected for the factors $\mathbf{Q}^2$ and 1/energy in the scattering cross-section. The calculated structure factors were normalized to the corrected integrated intensity of the TA phonon at **Q** = (0.2, 0.2, 4) where the structure factor is practically model-independent. Obviously, the calculated structure factors are borne out by experiment quite well, giving credence to the calculated eigenvector. We note that in this case, we did not observe any change of the structure factor with temperature.

As discussed in section IV.4, the phonon pattern of the TA mode at **q** = (0.5, 0.5, 0) predicted by DFPT and a BvK model differ from each other quite significantly (Fig. 8c) which is reflected in different structure factors. As can be seen from Tab. 2, DFPT is in much better agreement with experiment than the BvK model. We conclude that the large elongations of the B and C atoms calculated in DFPT – an apparently necessary ingredient for a strong EPC - are realistic.

### VI. Discussion and Conclusions

We have presented a combined theoretical and experimental investigation of the lattice dynamical properties of the phonon-mediated superconductor $YNi_2B_2C$. We found a very good agreement between theory and experiment not only for phonon frequencies but also for phonon line widths, i.e. for the predicted EPC strength, and that over the whole range of phonon energies up to 160 meV. In particular, our DFPT calculations with high resolution in energy-momentum space allowed us to find previously unknown phonon anomalies. We showed in very detailed investigations of some phonon modes having an exceptionally strong coupling that a realistic description of the wave vector dependent EPC rests upon a precise prediction of phonon eigenvectors. Here, large amplitudes of the light elements B and C play a crucial role, because B and C related electronic states have a large contribution to electronic density of states at the Fermi energy. We were able to experimentally verify the relatively large motions of the light elements predicted by theory even in some low energy acoustic phonon branches, which are usually dominated by movements of the heavy ions. This makes it understandable that DFPT predicts two thirds of the total EPC constant λ to come from phonons with energies less than 30 meV. Thus, our inelastic neutron scattering results demonstrate that DFPT accurately describes the lattice dynamical properties of $YNi_2B_2C$ and gives a very good account of EPC in this compound. Moreover, by varying the smearing parameter σ in the calculation of the electronic structure DFPT can even model the temperature dependence of energies and line widths of phonons with strong EPC, i.e. to relate the temperature dependence of phonon properties to the smearing of the Fermi edge. The caveat is that this works only on the qualitative level and the calculated temperature scale for phonon renormalization is much higher than the observed one. Understanding the origin of this discrepancy is subject of future work.

In addition to a powerful tool for simulating and predicting results of phonon measurements, DFPT can also be used to understand the details of electron-phonon coupling mechanisms in materials, because it allows calculations of electron-phonon coupling for electronic transitions band-by-band (for both interband and intraband transitions). Thus it is possible to learn which bands are responsible for which phonon anomalies (see for example [28,37]. Although a detailed investigation of this type was outside the scope of the present work, we can make a few steps in this direction based on the calculations and experimental results presented here.

Kohn anomalies received a lot of attention recently in the context of CDW formation. Two distinct mechanisms that lead to strong phonon softening on cooling emerged. In some compounds, such as 1D conductors, this softening is related to the nesting of the Fermi surface [38]. In others, such as $2H-NbSe_2$, it is related to **q**-dependence of electron-phonon coupling matrix elements [27,39]. In the case of Fermi surface nesting, smearing of the Fermi surface at finite temperature smears out the singularity in the joint density of states, which reduces the phonon softening at the wave vector of the singularity [26]. No such effect is expected when no nesting occurs and temperature should not have a strong effect on Kohn anomalies in the case when electron-phonon matrix elements are responsible. The observed phonon softening on cooling in such materials is then explained by the phonon sitting in a potential whose bottom is strongly anharmonic. As the temperature drops, the phonon increasingly feels this anharmonicity and, hence, softens and broadens. CDW transition occurs in the case of a double-well potential with the CDW transition temperature corresponding to the height of the barrier between the wells [39].

The Fermi surface in YNi$_2$B$_2$C is not nested in the sense that there are no singularities in the joint density of states that are strong enough to be exclusively responsible for the observed temperature dependence of the phonons. (Fig. 3) On the other hand, as mentioned above, the observed redistribution of the phonon lineshapes below T$_c$ is possible only if the linewidth is electron-phonon in nature, not anharmonic. So YNi$_2$B$_2$C has temperature-dependent phonon anomalies without nesting and without anharmonicity. So how to explain this apparent paradox? Our results point in the direction of a "hybrid" approach. If the Fermi surface is nested for a particular k$_z$ and the matrix elements for a particular phonon are exceptionally strong at that k$_z$, then the phonon effectively "sees" a nested 2D Fermi surface even when the total 3D joint density of states does not have a singularity. Thus we propose that, for example, the phonon at (0.56,0,0) couples strongly to the nested part of the Fermi surface with k$_z$=0.5 shown in Fig. 3a and this nesting in 2D is responsible for the Kohn anomaly at this wave vector. Going beyond the speculative discussion above requires considerable amount of additional work that will be carried out in the future.

**Acknowledgements**
F.W. was supported by the young investigator group of the Helmholtz Society under Contract No. VH-NG-840. D.R. was supported by the DOE,Office of Basic Energy Sciences,Office of Science, under Contract No. DE-SC0006939.